\documentclass[12pt,preprint]{aastex}
\usepackage{amsmath}
\usepackage{amsfonts}
\usepackage{amssymb}
\usepackage{hyperref}

\slugcomment{Not to appear in Nonlearned J., 45.}

\begin{document}

\title{Theoretical deduction of the Hubble law beginning with a \\
MoND theory in context of the $\Lambda$FRW-Cosmology}

\shorttitle{Deduction of the Hubble law from a MoND theory \\
in context of the $\Lambda$FRW-Cosmology}

\author{Nelson Falcon
and Andr\'{e}s J. Aguirre G.}

\affil{Laboratory of Physics of the Atmosphere and the Outer Space, University
of Carabobo, Valencia, Venezuela;
\href{mailto:nelsonfalconv@gmail.com}{nelsonfalconv@gmail.com},
\href{mailto:aaguirre3@uc.edu.ve}{aaguirre3@uc.edu.ve}}

\begin{abstract}
We deducted the Hubble law and the age of the Universe, through the
introduction of the Inverse Yukawa Field (IYF), as a non-local additive
complement of the Newtonian gravitation (Modified Newtonian Dynamics). As result we connected the
dynamics of astronomical objects at great scale with the
Friedmann-Robertson-Walker ($\Lambda$FRW) model. From the corresponding
formalism, the Hubble law can be expressed as $v=(4\pi \lbrack G]/c)r$,
which was derivated by evaluating the IYF force at distances much greater
than 50Mpc, giving a maximum value for the expansion rate of the universe of 
$H_0^\text{(m\'ax)} \simeq 86,31$km s$^{-1}$ Mpc$^{-1}$, consistent with the
observational data of 392 astronomical objects from NASA/IPAC Extragalactic
Database (NED). This additional field (IYF) provides a simple interpretation
of dark energy as the action  a large scale of baryonic matter. Additionally,
we calculated the age of the universe as $11$Gyr, in agreement with recent
measurements of the age of the white dwarfs in the solar neighborhood.
\end{abstract}


\section{Introduction}

The idea of ​​a model for a universe in continuous and constant expansion,
emerged from the pioneering work of Hubble, Slipher and Humason \citep{hub29}.
This dynamic description of the universe began with early studies on
relativistic cosmology \citep{lem27} and is the foundation of the Big Bang
theory, which explicitly uses the so-called Hubble law.

Hubble law was empirically proposed by \cite{hub29}, who noted a roughly
linear relation between velocities and distances among nebulae, and saw that
the relation appears to dominate the distribution of velocities. The
mathematical expression for this relation proposed by Hubble, so-called Hubble
law, is usually written as
\begin{equation}
 v = H_0 r,
\label{eq usual hubble law}
\end{equation}
where $v$ (in units of km s$^{-1}$) is the recessional velocity of a given
astronomical object, which distance from the Earth, $r$, is measured in Mpc,
and $H_0$ is the Hubble constant (in km s$^{-1}$ Mpc$^{-1}$), that can be
alternatively written as $H_0 = 100h \text{km s$^{-1}$ Mpc$^{-1}$}$, where
$h$ is the adimensional Hubble parameter and takes values from 0 to 1.

Since the discovery of the accelerating expansion of the universe through the
study of high redshift supernovae by \cite{rie98} and \cite{per99}, the
current cosmological model uses the Hubble law together with Friedmann
equations as the basis of the Standard Model of Big Bang cosmology.
Friedmann equations constitute the solutions of the Einstein field equations
of the General Theory of Relativity for the Friedmann-Robertson-Walker (FRW)
metric under the addicional assumption of an isotropic and homogeneous
universe at large scales (Cosmological Principle).

Although Hubble law represents the first observational test of the expansion
of the universe, and today supports the actual cosmological model
\citep{fre10}, it have not been theoretically deducted, so many hypotheses
have arisen to this end, and even more new
theories have emerged as alternative for the velocity-distance law. Among the
alternatives is \cite{bro62}, who determined that Hubble law is a linear
approximation of a more general exponential law, but it was conceived for a de
Sitter universe with no matter. \cite{seg93} by studying IRAS data proposed a
square law, as given by Lundmark, however \cite{str93} reviewed Segal's
research and studied IRAS data too and determined that observations actually
support Hubble law, the same result was obtained for galaxies from CfA and
ESO/LV \citep{cho95}. \cite{pas00} determined a generalized Hubble law which
introduces two addicional terms to the usual Hubble law produced by the
angular expansion, but this conception implies an anisotropic universe in
conflicts with the Cosmological Principle. At this point, Hubble law has
remained unalterable, and therefore the latest theories seem to look for
deriving the velocity-distance law as was proposed by Hubble, namely, they
look for a theoretical deduction of the Hubble law. \cite{jia05} derived the
Hubble law under a hypothesis that eliminates the need for dark energy,
nevertheless he used a non-conventional form of the FRW metric with the time
defined as relative to some hypothetical time where the line element was or
will be the Minkowskian, which has not been found by observations.
\cite{sor09} proposed that Hubble law, as result of an expanding universe, is
really a working hypothesis, instead he considered the hypothesis proposed by
Zwicky of the tired-light, but nowadays it is well known that this
theory is not supported by observations, in fact it does not explain the
anisotropies in the CMB. Recently, \cite{san14} opted for a non-standard form
of the Hubble law, assuming a new definition of the redshift based in
frequencies rather than wavelenght, establishing a new paradigm for the
spectroscopy.

One of the biggest problems in the Big Bang cosmology, closely linked to the
expansion of the Universe and the Hubble law, is the evidence of the
accelerated expansion of the Universe, commonly referred
as dark energy, whose understanding is still far from complete. Also, the
inconsistency between the observed average density of matter and the density
required for flatness of the universe, a problem known as the missing mass,
has become the paradigm of the hypothetical non-baryonic dark matter. This
discrepancy between the astronomical observations of the density of matter and
expected in $\Lambda$FRW model in the Big Bang theory, has prevailed in the
last years. An alternative to the paradigm of non-baryonic dark matter, is the
theory of Modified Newtonian Dynamics (MoND), which involve changes in the
Newton's law of gravitation (inverse square law).

In this sense, one possibility to solve both problems: dark matter and dark
energy, is the non-local gravitation recently proposed by \cite{fal13}, which
basically is a MoND theory. According to which the force of gravitation would
be the result of two fields generated by the ordinary baryonic matter, a first
term as Newton law of inverse square, and an addicional long-range term.

The inclusion of this second term in the force of gravity, consistent with
E\"otv\"os-like experiments, can reconcile the $\Lambda$FRW model with
observables of the Big Bang, without the paradigm of non-baryonic dark matter.
Additionally, gives an explanation for dark energy, and allows
theoretically deduce the Hubble law.

In this paper, we will show that the Hubble law derivates from the MoND theory
proposed by \cite{fal13} in a natural way through the corresponding condition
of cosmological scales. To this end, in section \ref{sec IYF} we will review
the paper of Falc\'on emphasizing the repulsive behavior of the non-local
gravitational field at large scales, giving a starting point for deducting the
Hubble law. The theoretical deduction of the Hubble law and even an analytical
determination of the Hubble constant will be given in sections \ref{sec
deduction}. In section \ref{sec test}, we will contrast the determined Hubble
constant with the observational data of 392 objects selected from the
NASA/IPAC Extragalactic Database (NED), also a brief discussion about the
cosmic age problem is given. Finally, the conclusions are given in
section \ref{sec conclusions}.

\section{MoND with non-local gravitational term}
\label{sec IYF}

Current Big Bang cosmology assumes Newtonian gravitation as the only
fundamental force at astronomical scales, giving a complete determination of
the dynamics of the universe. However, from this idea we find serious
difficulties to describe the behavior of the Universe: (1) galaxy rotation
curves are not explained without the inclusion of non-baryonic dark matter,
whose fundamental nature and properties are completely unknown, (2) into the
rich galaxy clusters, the observed mass of stars and the gas mass inferred
from the X-ray diffuse emission is significantly less than that required to
hold these systems gravitationally stable, and (3) the
accelerated expansion of the universe violates our understanding about how
gravity works at cosmological scales (see \cite{fal13} for details).

The simplest way for modeling the accelerated cosmic expansion is by
introducing a cosmological constant into the Einstein's field equations so it
can representate an hypothetical negative pressure of the vacuum of space,
also called dark energy. However this is given as a disconnected idea from the
dynamics of the astronomical objects, which is limited to the Newton's law of
gravitation.

\begin{figure}[t]
 \centering
 \includegraphics[scale=.6]{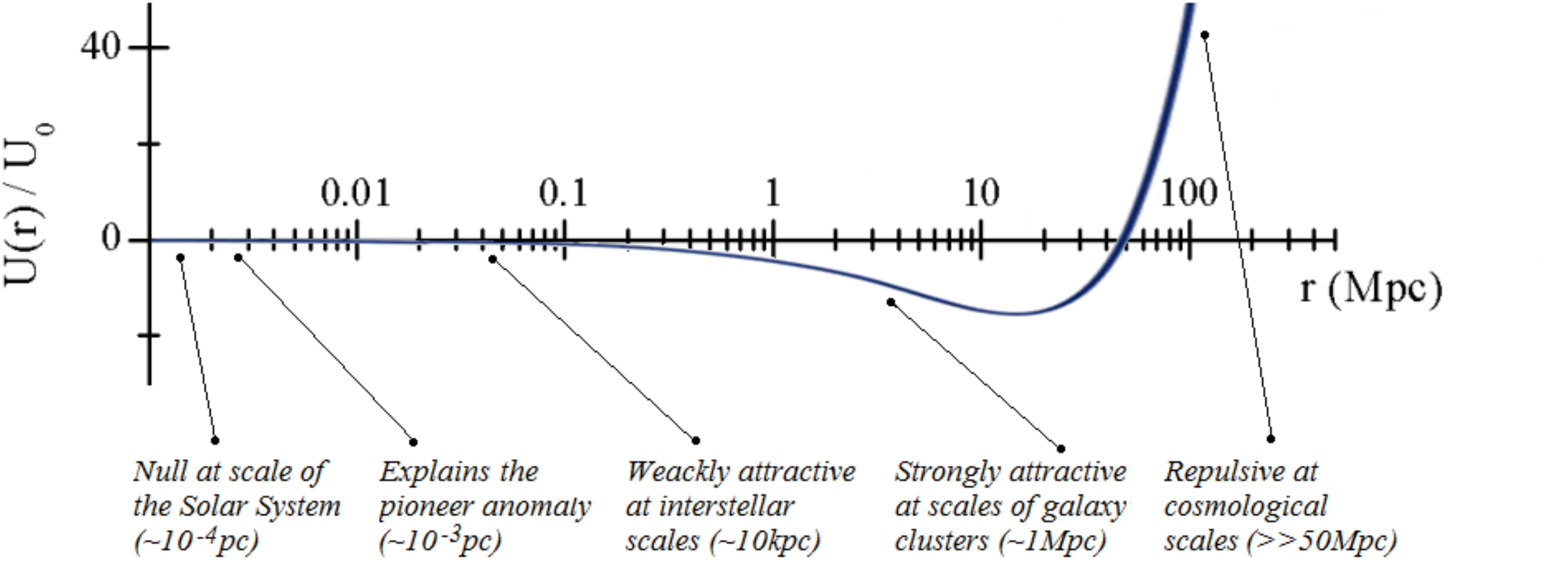}
 \caption{IYF potential per unit of mass as function of the distance between
 objects gravitationally bounded (see \cite{fal13} for details).}
 \label{fig IYF potential}
\end{figure}

While Newtonian gravitation (inverse square law) has been highly supported by
laboratory experiments and satellites, there is no experimental evidence to
confirm its validity beyond the Solar System \citep{gun05}. That is why it has
raised the Modified Newtonian Dynamics (MoND) theories such as proposed by
\cite{mil83}, that solves the galaxy rotation problem prescinding from
non-baryonic dark matter. Following this line, \cite{fal11,fal13} proposed a
modification of the Newtonian gravitation by adding a non-local term that
contains Milgrow's theory as a particular case and establishes a possible
connection for the dynamics at large scale and FRW formalism. This additional
term was constructed by the specular reflection of the potential of Yukawa, so
that we decided to named it: Inverse Yukawa Field (IYF). This
interaction is given by the baryonic matter (as the
Newtonian gravity), and shows a null contribution at scale of the Solar System
($\sim 10^{-4}$pc), in agreement with measurements on Earth, weackly atractive
at interstellar distances ($\sim 10$kpc), consistent with MoND theory (as a
solution of the galaxy rotation problem), strongly attractive at scales of
galaxy clusters ($\sim 1$Mpc), in accordance with Abell radius, and repulsive
at cosmological scales ($\gg50$Mpc), in agreemente with the expansion of the
universe (see Figure \ref{fig IYF potential}). This interaction has a potential
per unit of mass of the form
\begin{equation}
U\left( r\right) \equiv U_{0}\left( r-r_{0}\right) e^{-\alpha /r},
\label{eq IYF potential}
\end{equation}
where $U_{0}=U_{0}\left( M\right) $ is the magnitude of the potential (in
units of N kg$^{-1}$) as a function of the baryonic matter, $\alpha \sim
2,5h^{-1}$Mpc and $r_{0}\sim 50h^{-1}$Mpc are constants.

Then, the proposed modification considers the contribution of both the
Newtonian and the non-local gravitational field, so that the dynamics at all
scales is determined by the force per unit of mass as
\begin{equation}
 F(r) = G \frac{M}{r^2} - \frac{U_{0}\left( M\right) }{r^{2}}
e^{-\alpha /r}\left[ r^{2}+\alpha \left( r-r_{0}\right) \right] ,
\label{eq force}
\end{equation}
where it is important to note that there is a dependence on the baryonic
matter only.

In particular, a zero contribution of the non-local term can be verified at
distances below $10^{-4}$pc, in agreement with measurements on Earth as
E\"otv\"os-like experiments. However, a measurable contribution can be
observed at 45AU, indeed the IYF provides a sunward acceleration of the order
of $10^{-11}$m s$^{-2}$ consistent with acceleration presented by the pioneer
spacecraft \citep{tur10}. On the other hand, at scales of tens of kiloparsec,
the Newtonian contribution can be neglected and the IYF term shows a MoND-like
behavior of the form
\begin{equation}
F \left( r\ll r_{0}\right) \simeq \left( \frac{U_0 \left( M \right)
r_0 }{2}\right) \frac{1}{r},
\end{equation}
solving the galaxy rotation problem. Also, the non-local IYF, evaluated in the
Abell radius ($r \sim 1,2$Mpc), provides an additional force, two hundred and
fifty times greater than the Newton's force, so it could solve the missing
mass problem in galaxy clusters first identified by Zwicky.

From Figure \ref{fig IYF potential}, it is clear that IYF potential gives
a constant repulsive force at cosmological scales ($\gg50$Mpc) as
\begin{equation}
F \left( r \gg 50 \text{Mpc} \right) \simeq U_0 \left( M \right) ,
\end{equation}
providing an asymptotic cosmic acceleration, consistent with the observations.
This opens the possibility to describe the behavior of the cosmological
constant by setting it as a dynamical term with the form
$\Lambda \equiv \Lambda (r) \propto U(r)$, giving a link between the dynamics
of astronomical objects, due the Newtonian and IYF force, with Friedmann
equations, which are only modified by the introduction of the dynamism of
$\Lambda$ as
\begin{equation}
\left( \frac{\dot{R}\left( t\right) }{R\left( t\right) }\right) ^{2}+\frac{%
kc^{2}}{R^{2}\left( t\right) }=\frac{8\pi G}{3}\rho +\frac{\Lambda \left(
r\right) c^{2}}{3}
\label{eq friedmann 1}
\end{equation}
\begin{equation}
2\frac{\ddot{R}\left( t\right) }{R\left( t\right) }+\left( \frac{\dot{R}%
\left( t\right) }{R\left( t\right) }\right) ^{2}+\frac{kc^{2}}{R^{2}\left(
t\right) }=-\frac{8\pi G}{c^{2}}P+\Lambda \left( r\right) c^{2},
\label{eq friedmann 2}
\end{equation}
where the dot denotes the time derivate of the scale factor $R(t)$,
$k=-1,0,+1$ is the scalar curvature for a open, flat and closed universe
respectively, $c$ is the speed of light, $G$ is the gravitational constant,
$\rho$ is the total mass-energy density, and $P$ the pressure.

The introduction of the non-zero contribution of the cosmological
constant brings a modification to the usual form of the matter density
parameter, $\Omega_m$, in terms of the energy of the IYF, $\Omega_\text{IYF} \simeq 8,1$
\citep{fal13}. Then, Eq. \eqref{eq friedmann 1} is now
\begin{equation}
\frac{kc^{2}}{R^{2}\left( t\right) }=H_{0}^{2}\left[ \Omega _{m}\left(
1+\Omega _{\text{IYF}}\right) + \Omega _\Lambda - 1 \right] ,
\end{equation}
where the dark energy density parameter, $\Omega _\Lambda$, is defined as
usual. Hence, the flatness condition ($k = 0$) is fulfilled by Friedmann
equation without the assumption of non-baryonic dark matter.

For a complete interpretation of the behavior of the IYF potential and details
about the cosmological consequences by adding the IYF to Newtonian dynamics
and to FRW cosmology see \cite{fal13}. Finally, the repulsive behavior of this
non-local term provides an starting point for studying the dynamics at large
scale, and therefore for deducting theoretically the Hubble law.

\bigskip

\section{Theoretical deduction of Hubble law}
\label{sec deduction}

A numerical value for $U_{0}$ can be found by studying the gravitational
Poisson equation, noting that the IYF potential must satisfy this equation.

Usually, the Poisson equation is written for the Newtonian gravitational
case as $\nabla^2 \Phi _{\text{N}}=4\pi G\rho $, however since the
introduction of the new interaction we must add a scalar field
corresponding to the IYF, so $\nabla^2 \Phi _{\text{N}} \rightarrow \nabla^2
\Phi _{\text{N}} + \nabla^2  \Phi _{\text{IYF}}$, but because we are
evaluating the asymptotic limit of
cosmological scales, the Newtonian contribution is not important. Thus, the
Poisson equation with IYF for a spatial matter distribution, $\rho \left( r
\right) $, is
\begin{equation}
\nabla ^{2}U\left( r\right) =4\pi G\rho \left( r\right) ,  \label{eq Poisson}
\end{equation}
with $U\left( r\right) $ given by Eq. \eqref{eq IYF potential}. Therefore,
calculating the laplancian operator with spherical symmetry, and observing
that the resulting function and the density are lineraly dependent for $r \gg
50$Mpc (in a mathematical sense), we have that the constants $U_{0}$ and
$4\pi G$ must satisfy the equality $U_{0}=4\pi G$, where the units are
specified as 
\begin{equation}
U_0 = 4 \pi G\left( \text{kg m}^{-2}\right) \simeq 8,39 \times 10^{-10}
\text{N kg}^{-1} .
\label{eq U0}
\end{equation}
and the same for the matter density
\begin{equation}
\rho \left( r\right) =\frac{e^{-\alpha /r}}{r^{4}}\left[ 2r^{2}\left(
r+\alpha \right) +\alpha ^{2}\left( r-r_{0}\right) \right] \left(
\text{kg m}^{-2}\right) ,
\end{equation}
taking into account that this equality works for $r$'s much greater than
$50$Mpc, so that the Newtonian contribution is null.

Here, we note that the obtained magnitude, $U_0$, for the IYF potential gives
a maximum value, as result of evaluating the behavior of the baryonic matter
density in the asymptotic case of cosmological scales.

On the other hand, in section \ref{sec IYF} we saw that the IYF, as a
non-local term, shows a repulsive behavior at cosmological scales ($\gg
50$Mpc), providing an asymptotic cosmic acceleration
in accordance with accelerated expansion of the universe. This allows a
theoretical deduction of the Hubble law, as a lineal proportionality between
recessional velocities and distances.

Consider a particle (galaxy, galaxy cluster, nebulae, etc.) with nonzero
rest mass under the influence of the IYF force. The contribution of the
Newtonian gravitational force is not important at cosmological escales (i.e.
at 1Mpc, $F_{\text{Newton}}\propto r^{-2}\sim 10^{-49}$N at least). Thus,
the equation of motion is only given by the force per unit of mass of the
IYF. Additionally, since the IYF is conservative, we can write 
\begin{equation}
\frac{d^{2}r}{dt^{2}}=\frac{dU}{dr}=\frac{U_{0}\left( M\right) }{r^{2}}
e^{-\alpha /r}\left[ r^{2}+\alpha \left( r-r_{0}\right) \right] ,
\label{eq eq of motion}
\end{equation}
where $U_{0}=U_{0}\left( M\right) $ depending of the baryonic matter, $M$,
that causes the field.

Then, it is possible to obtain an expresion of the velocity by integrating
Eq. \eqref{eq eq of motion} as $v=\int \left( d^{2}r/dt^{2}\right) dt$.
Here, the time interval $dt$ is measured through the photons giving the
recessional velocity of the particle. Therefore,
\begin{equation}
v=\int \frac{dU}{dr}\frac{dr}{c}\Longrightarrow v=\frac{1}{c}U+\text{const},
\label{eq integral}
\end{equation}
with $v$ as the time derivate of the comoving distance, $r$. Then, the
velocity is proportional to the IYF potential. Actually, because the Hubble
flow is observed at cosmological distances, we should evaluate the IYF
potential at $r\gg 50$Mpc, so that $r_0/r \simeq 0$ and $e^{-\alpha /r}
\simeq 1$. Hence, from Eq. \eqref{eq IYF potential}, we obtain
\begin{equation}
v=\frac{U_{0}}{c}r,
\label{eq My Hubble law}
\end{equation}
where without loss of generality we assumed the initial condition $v = 0$ at
$r=0$, resulting in a null integration constant. Here, from
analogy with the Hubble law we find the Hubble constant as $U_0 /c$,
where the magnitude of the IYF potential is given by Eq. \eqref{eq U0}. Even
more, because we are
evaluating the asymptotic limit of cosmological scales in the IYF force, we
can determine the limit value of this proporcionality constant as
\begin{equation}
H_0^\text{(m\'ax)} = \frac{4\pi G}{c}\left( \text{kg m}^{-2}\right) \simeq
86,31 \text{km\ s$^{-1}$\ Mpc$^{-1}$} ,
\label{eq H max}
\end{equation}
so that the Hubble law can be written as
\begin{equation}
v = H_0^\text{(m\'ax)} r.
\label{eq hubble law}
\end{equation}

Note that Eq. \eqref{eq hubble law} basically is equal to Eq. \eqref{eq usual
hubble law}, establishing a linear relation between recessional velocities and
distances for a given particle (galaxy, cluster of galaxy, nebulae, etc.),
just under the assumption of cosmological scales, in agreement with the
current cosmological model. Additionally, the limit value of the linearly
constant gives the maximum expansion rate of the universe, again as product of
study distances much greater than 50Mpc.

In the next section, we will test the $H_0^\text{(m\'ax)}$ value with the
observational data from NED, under criteria that allow stuying the
velocity-distance relation at large scale.

\section{Observational test and discussions}
\label{sec test}

Although the first determination of the Hubble constant was $H_0 = 500$km
s$^{-1}$ Mpc$^{-1}$ \citep{hub29}, today it is
well known that this proporcionality constant takes values less than 100km
s$^{-1}$ Mpc$^{-1}$. In fact, \cite{san58} gave the first reasonable estimated
of the Hubble constant by studying Cepheids, he obtained that $H_0$ is about
75km s$^{-1}$ Mpc$^{-1}$. Four decades later, \cite{fre01} studied objects
over the range of about 60-400Mpc, using Cepheids, and determined that $H_0
= 72 \pm 8$km s$^{-1}$ Mpc$^{-1}$. \cite{bon06} studied galaxies with
redshift between 0,14 and 0,89, obtaining that the Hubble constant is
$77,6_{-12,5}^{+14,9}$km s$^{-1}$ Mpc$^{-1}$. After nine years of recording
and analysis of the CMB data from WMAP, \cite{ben12} calculated that $H_0 =
69,32 \pm 0,80$km s$^{-1}$ Mpc$^{-1}$. The latest value of the Hubble constant
was determined by \cite{ade13}, who studied the CMB through Planck satellite,
where the data fitted $H_0 = 67,15\pm 1,20$km s$^{-1}$ Mpc$^{-1}$, being the
value accepted today.

In this section, we will comparate our value for the Hubble constant, of
$H_0^\text{(m\'ax)} \simeq 86,31$km s$^{-1}$ Mpc$^{-1}$, with the
observational data provided by the NASA/IPAC Extragalactic Database (NED).

In order to varify the Hubble law and our value for the Hubble constant, we
will use the primitive technique used by \cite{hub29}, which consists
in plotting the observational measurements of the velocity (via redshift)
and the distance of a set of objects such as galaxies, quasars, radio
sources, X-ray sources, infrared sources, etc. For this end, we considered the
Master List of Redshift-Independent Extragalactic Distances of 15339 galaxies
provided by NED (Version 9.2.0). The observational measurements were filtered
by: (1) recent measurements (year of publication from 2009), (2) distant
objects ($r\gg 50$Mpc), (3) redshift from $0,0167$ to $0,33$, and (4) accurate
measurements (with maximum error of 0,5\%). As result, the list was reduced to
392 objects (the complete list of the 392 objects can found on
\url{https://db.tt/vwlVdhVM}). Here, we must clarify that in
order to filter
errors by peculiar motions we used redshifts above 0,0167, so errors are
under 6\% \citep{fre01}, and due the theoretical assumption of $v=cz$ we set
redshifts below 0,33, so that Lorentz factor is equally under 6\% and the
relativistic effects are neglected.

\begin{figure}[t]
 \centering
 \includegraphics[scale=5]{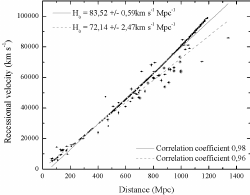}
 \caption{Hubble diagram for objects at 50-1400Mpc (data from NED).}
 \label{fig Large scale}
\end{figure}

A Hubble diagram for the 392 galaxies, in a range of 50-1400Mpc, is shown in
Figure \ref{fig Large scale}. Through a linear fit we found that $H_0 \simeq
83,56\pm 0,59$km s$^{-1}$ Mpc$^{-1}$ (solid line), so that the determined
maximum expansion rate of $H_0^\text{(m\'ax)}=86,31$km s$^{-1}$ Mpc$^{-1}$
disagrees about 3\% only. Nevertheless, because our prediction works at the
limit of cosmological scales, we would hope that observational measurements
shows an asymptotic behavior to an expansion rate of $86,31$km s$^{-1}$
Mpc$^{-1}$ at distances even greater than 1500Mpc. Actually, in Figure
\ref{fig Large scale}, we can note that observational data slightly suggest an
upper slope for distances above 1000Mpc.

In Figure \ref{fig Large scale}, we additionally note that the data suggest a
lower slope for distances lower than 500Mpc. Actually, a linear fit at that
scales gives a Hubble constant of $H_{0}=72,14\pm 2,47$km s$^{-1}$ Mpc$^{-1}$
(dashed line), which agrees with the value given by
\cite{fre01} of $H_{0}=72\pm 8$km s$^{-1}$ Mpc$^{-1}$ (via Cepheid variables
applied over the range of about 60-400 Mpc) and \cite{bla01} of
$H_{0}=72\pm 15$km s$^{-1}$ Mpc$^{-1}$ (via SBF with range of applicability
until 125 Mpc). Clearly, from the used method this consistence is expected.

On the other hand, an additional result can be obtained through the determined
Hubble constant: the age of the universe, $\tau$, which can be calculated as
\begin{equation}
\tau = H_{0}^{-1}\int\nolimits_{0}^{\infty }\left[ \left(
1+z\right) ^{3}\Omega _{m}\left( 1+\Omega _{\text{IYF}}\right) +\Omega
_{\Lambda }\right] ^{-1/2}\frac{dz}{1+z}
\label{eq age integral}
\end{equation}
\citep{fal13}, where $\Omega_m$ is the density parameter of matter but only
including baryonic matter, $\Omega_\text{IYF}$ is a density parameter emerged
from the contribution of the IYF, $\Omega_\Lambda$ is the cosmological density
parameter, and $z$ is the usual redshift. Here, we note that Eq. \eqref{eq
age integral} is reduced to the conventional form, in the $\Lambda$FRW model
with $k=0$, when the IYF is zero.

Numerical integration of Eq. \eqref{eq age integral}, for $H_0^\text{(m\'ax)}
=86,31$km s$^{-1}$ Mpc$^{-1}$, $\Omega _{m}=0,03$ \citep{fre03},
$\Omega_{\text{IYF}} = 8,80$ and $\Omega _{\Lambda }=0,71$ \citep{fal13},
gives an age of the universe of $\tau \simeq 11 $Gyr, in agreement with the
age of the white dwarfs in the solar neighborhood
\citep{tre14} and, from the Copernican Principle, with the age of the
universe. Here, we must say that there exist astronomical objects with ages
greater than 11Gyr, i.e., B495 is $14,54$Gyr, B024 is $15,25$Gyr and
B050 is $16,00$Gyr \citep{wan10}, however the determination of the age of the
oldest globular clusters, via HR diagram, introduces intrinsic errors of about
25\%, in which case they would be consistent with an age of 11Gyr.

\section{Conclusions}
\label{sec conclusions}

The inclusion of a long-range component in the law of gravitation allows
linking the Hubble law with the dynamics of the large-scale Universe.
Particularly, if the non-locality of gravitation is included through a
potential as shown here, Yukawa Inverso type, we can connect the dark energy
with cosmological constant and derive from there the Hubble law, consistent
with the formalism of the Big Bang, and astronomical observations. All this
without resorting to the paradigm of non-baryonic dark matter, or an ``exotic
physics''.

The inclusion of a long-range component in the law of gravitation, through an
inverse potential Yukawa-like, represent the collective contribution of the
gravitational effects of large-scale, on the order of tens of megaparsec
caused by ordinary baryonic matter. In this sense, the IYF explicitly includes
the Mach principle in the formalism of FRW cosmology, as pretended by Einstein
with the Theory of General Relativity.

The prescription of the Hubble constant in terms of the fundamental constants,
as in Eq. \eqref{eq H max}, appears to correspond to the observational data
for distant objects, whose distance and redshift are independently known; as
we can see in Figure \ref{fig Large scale}. Note that the Hubble constant is
not measured directly by the WMAP and Planck satellites, but rather its value
is inferred from the power spectrum of the cosmic background radiation (CMB)
together with other cosmological variables through multiple statistics
correlation, or maximum likelihood.

For the nearest objects, with distances less than a hundred megaparsec, the
Hubble constant would seem less than true valor, because in these ranges, the
contribution of the IYF field is less, as was shown in Figure \ref{fig IYF
potential}.

A current Hubble constant ($H_0$) of higher value, such as $H_0^\text{(m\'ax)}
\simeq 83,56 \pm 0,59$km s$^{-1}$ Mpc$^{-1}$, implies a more recent age for
the universe, but still, this value is surprisingly similar to that inferred
for the age of the oldest white dwarfs in the Milky Way. Obviously the Milky
Way would have to be as old as the universe itself under the Copernican
Principle, which is the very foundation of the Big Bang theory.

\acknowledgments

This research has made use of the NASA/IPAC Extragalactic Database (NED) which
is operated by the Jet Propulsion Laboratory, California Institute of
Technology, under contract with the National Aeronautics and Space
Administration.


\begin{thebibliography}

\bibitem[Ade et al.(2013)]{ade13} Ade, P. A. R. et al. 2013, arXiv:1303.5062v1

\bibitem[Bennett et al.(2012)]{ben12} Bennett, C. et al. 2012, \apj, 208, 1.

\bibitem[Blakeslee et al.(2001)]{bla01} Blakeslee, J. et al. 2001, \mnras, 327, 1004.

\bibitem[Bonamente et al.(2006)]{bon06} Bonamente, M. et al. 2006, \apj, 642, 25.

\bibitem[Browne(1962)]{bro62} Browne, P. F. 1962, \nat, 193, 1019.

\bibitem[Choloniewski(1995)]{cho95} Choloniewski, J. 1995, arXiv:astro-ph/9504035

\bibitem[Falc\'{o}n(2011)]{fal11} Falc\'{o}n, N. 2011, \rmxaa, 40, 11.

\bibitem[Falc\'{o}n(2013)]{fal13} Falc\'{o}n, N. 2013, J. Mod. Phys., 319, 10.

\bibitem[Freedman et al.(2001)]{fre01} Freedman, W. et al. 2001, \apj, 553, 47.

\bibitem[Freedman \& Turner(2003)]{fre03} Freedman, W. \& Turner M. 2003, Rev. Mod. Phys, 75, 1433.

\bibitem[Freedman \& Madore(2010)]{fre10} Freedman, W. \& Madore, B. 2010, \araa, 48, 673.

\bibitem[Gundlach(2005)]{gun05} Gundlach, J. H. 2005, New J. Phys., 7, 205.

\bibitem[Hubble(1929)]{hub29} Hubble, E. 1929, Proc. Natl. Acad. Sci. USA, 15, 168.

\bibitem[Lema\^{\i}tre(1927)]{lem27} Lema\^{\i}tre, G. 1927, Ann. Soc. Sci. de Bruxelles, 47, 49.

\bibitem[Liu(2005)]{jia05} Liu, Jian-Miin 2005, arXiv:physics/0507018

\bibitem[Milgrom(1983)]{mil83} Milgrom, M. 1983, \apj, 270, 371.

\bibitem[Pascual-S\'{a}nchez(2000)]{pas00} Pascual-S\'{a}nchez, J.-F. 2000, arXiv:gr-qc/0010076

\bibitem[Perlmutter et al.(1999)]{per99} Perlmutter, S. et al. 1999, \apj, 517, 565.

\bibitem[Riess et al.(1998)]{rie98} Riess, A. et al. 1998, \aj, 116, 1009.

\bibitem[Sandage(1958)]{san58} Sandage, A. 1958, \apj, 127, 513.

\bibitem[Sanejouand(2014)]{san14} Sanejouand, Y.-H. 2014, arXiv:1401.2919

\bibitem[Segal et al.(1993)]{seg93} Segal, I. E. et al. 1993, \apj, 411, 465.

\bibitem[Sorrell(2009)]{sor09} Sorrell, W. H. 2009, Astrophys. Space Sci., 323, 205.

\bibitem[Strauss \& Koranyi(1993)]{str93} Strauss, M. \& Koranyi, D. 1993, arXiv:astro-ph/9308028

\bibitem[Tremblay(2014)]{tre14} Tremblay, P. E. et al. 2014, arXiv:1406.5173v1

\bibitem[Turyshev \& Toth(2010)]{tur10} Turyshev, S. G. \& Toth, V. T. 2010, Living Rev. in Relativity, 13, 4.

\bibitem[Wang et al.(2010)]{wan10} Wang, S., Li, X. \& Li, M. 2010, \prd, 82, 103006.
\end{thebibliography}
\end{document}